\documentclass[conference]{ieeeconf}

\IEEEoverridecommandlockouts  
\usepackage[utf8]{inputenc} 
\usepackage[T1]{fontenc}    
\usepackage{hyperref}       
\usepackage{url}            
\usepackage{booktabs}   
\usepackage{amsmath}
\usepackage{amsfonts}
\usepackage{xcolor}
\usepackage{soul}
\usepackage{nicefrac}       
\usepackage{microtype}      
\usepackage{lipsum}
\usepackage{fancyhdr}       
\usepackage{graphicx}       
\graphicspath{{media/}}     
\usepackage{caption}
\usepackage{subcaption}
\usepackage{stmaryrd}

\usepackage{algorithm} 
\usepackage{algpseudocode} 

\usepackage[utf8]{inputenc} 
\usepackage[T1]{fontenc}    
\usepackage{hyperref}       
\usepackage{url}            
\usepackage{booktabs}   
\usepackage{amsmath}
\usepackage{amsthm}
\usepackage{amsfonts}
\usepackage{mdframed}
\usepackage{graphicx}
\usepackage{xcolor}
\usepackage{epstopdf}
\usepackage{amssymb}
\usepackage{matlab-prettifier}
\usepackage{siunitx}
\usepackage{soul}
\usepackage{nicefrac}       
\usepackage{microtype}      
\usepackage{lipsum}
\usepackage{fancyhdr}  
\usepackage[ragged]{sidecap}    

\usepackage{graphicx}       
\graphicspath{{media/}}     
\usepackage{caption}
\usepackage{subcaption}
\usepackage{stmaryrd}

\usepackage{algorithm} 

\usepackage{algpseudocode} \usepackage{cuted}
\algrenewcommand\algorithmicrequire{\textbf{Input:}}
\algrenewcommand\algorithmicensure{\textbf{Output:}}
\newtheoremstyle{thmv2}
{}
{.1em}
{\itshape}
{}
{\bfseries}
{.}
{.1em}
{}
\theoremstyle{thmv2}
\newenvironment{tightdisplay}{%
  \begingroup
  \setlength\abovedisplayskip{6pt}%
  \setlength\belowdisplayskip{6pt}%
  \setlength\abovedisplayshortskip{4pt}%
  \setlength\belowdisplayshortskip{4pt}%
}{\endgroup}

\usepackage{mathtools}
\usepackage{tikz}
\usetikzlibrary{shapes,arrows}
\usetikzlibrary{intersections}
\usetikzlibrary{automata,arrows.meta}
\usetikzlibrary{calc}
\usepackage{cancel}
\usepackage[mathscr]{eucal}
\usepackage{algpseudocode}

\newcommand{\until}{\mathbin{\sf U}}

\newcommand{\spaceR}{\mathbb{R}} 

\newcommand{\spaceX}{\mathbb{X}} 
\newcommand{\spaceS}{\mathbb{S}}

\newcommand{\spaceU}{\mathbb{U}} 

\newcommand{\spaceY}{\mathbb{Y}} 

\renewcommand{\qedsymbol}{}
\newcommand{\mdpM}{\mathbf{M}}

\newcommand{\s}[1]{s_{#1}} 
\newcommand{\AP}{\mathrm{AP}}
\newcommand{\letter}{l}
\newcommand{\word}{\boldsymbol{l}}

\newcommand{\DFA}{\mathcal A}

\newcommand{\history}{\mathcal H}

\newcommand{\subphi}{\alpha}

\newcommand{\valuemapping}{\eta}

\newcommand{\tensorV}{\mathscr{V}}
\newcommand{\qmapping}{\mathcal{L}_Q}

\newcommand{\expectation}{\mathbb{E}}
\newcommand{\indicator}{\boldsymbol{1}}

\newcommand{\op}{\mathbf{T}}

\newtheorem{definition}{Definition}
\newtheorem{example}{Example}

\newtheorem{lemma}{Lemma}
\newtheorem{theorem}{Theorem}

\newtheorem{proposition}{Proposition}

\begin{document}
\title{
Efficient control of multi-agent robust MDPs through tensor trees}
\author{Ruohan Wang$^{1}$, Giannis Delimpaltadakis$^{2}$, Luca Laurenti$^{2,3}$ and Sofie Haesaert$^{1}$ \thanks{This work is supported by the European project COVER under grant number 101086228 and by the Dutch Research Council (NWO) under grant OCENW.M.22.056.
$^{1}$Department of Electrical Engineering, Control
Systems Group, Eindhoven University of Technology, The Netherlands. $^{2}$Robust and Intelligent Autonomous Systems lab, Italian Institute of Artificial Intelligence (AI4I), Turin Italy. $^{3}$Delft Center for Systems and Control, Delft University of Technology, The Netherlands. Emails: r.wang2@tue.nl; ioannis.delimpaltadakis@ai4i.it; l.laurenti@tudelft.nl.; s.haesaert@tue.nl}
}

\maketitle
\begin{abstract}
This paper introduces a tensor-based framework for controller synthesis of stochastic multi-agent systems under temporal logic specifications that couple the agents, such as one agent reaching its goal only after another agent has reached its own. 
The agents are modeled as robust Markov decision processes (RMDPs), whose transition probabilities are only known to lie in an ambiguity set, for instance due to abstraction errors, learned models, or uncertain parameters. We show that the decoupled nature of the ambiguity sets allows one to decompose the computation of a lower bound on the satisfaction probabilities into per-agent factors, so that only the per-agent ambiguity sets need to be stored. This leads to improved memory complexity compared to monolithic approaches. We compare our approach with state-of-the-art approaches based on approximate simulation relations and with interval MDP abstractions. On a four-agent ordered reach-avoid benchmark, our approach certifies a lower bound twice as high as the simulation relation-based approach with 20 times less memory, and needs about 200 times less memory than the interval MDP abstraction, even at a resolution with about a million times more joint cells.

\end{abstract}


\section{Introduction}
\label{sec:introduction}
Teams of autonomous agents increasingly operate in domains where failures are costly or dangerous, such as warehouse logistics \cite{wurman2008coordinating} and disaster response with unmanned aerial vehicles \cite{erdelj2017help}. These agents may be required to complete tasks that constrain them jointly. For instance, robots carry packages in a convoy through a shared corridor, or drones reach their designated zones in a given order. Temporal logic specifications express such tasks formally, as requirements on the order and timing of events along the agents' behaviours \cite{baier2008principles,belta2017formal}. Agents' motion is subject to random disturbances, and the models used for control design are only approximate. Under uncertainty, a controller may fail the task, which can damage goods, delay a rescue operation, or endanger people nearby. Controllers therefore need guarantees that hold despite this uncertainty before they are deployed. Correct-by-design control synthesis provides such guarantees. It computes a controller together with a formal guarantee that the controlled system satisfies a temporal logic specification \cite{belta2017formal}. 
\par For stochastic systems, the guarantee is a certified lower bound on the probability that the specification is satisfied \cite{abate2008probabilistic,tkachev2017quantitative}. Arguably, the state-of-the-art approach to compute such guarantees is to rely on finite-model formal approximations of the continuous-state system, called abstractions \cite{lahijanian2015formal,mathiesen2025scalable}. 
However, the transition probabilities of such finite models are often known only within sets, due to abstraction errors \cite{lahijanian2015formal,haesaert2020robust}, estimation from data, or uncertain parameters \cite{suilen2024robust}. Robust Markov decision processes (RMDPs) model these uncertainties by an ambiguity set for each state and input: a set of transition distributions that contains those of the true system \cite{nilim2005robust,suilen2024robust}.
Interval MDPs are a widely used instance \cite{givan2000bounded,lahijanian2015formal, delimpaltadakis2023interval}. For multi-agent systems, the specification is evaluated on the joint behaviour of the agents.
Specification-decomposition-based methods split the team specification into per-agent tasks \cite{schillinger2018simultaneous}, or start from local tasks for each agent, possibly with synchronization \cite{tumova2016multi}. Compositional methods treat each agent separately against assumptions on other agents \cite{lavaei2022automated}. These methods do not yet address specifications that couple the agents. For such specifications, a common approach is dynamic programming on the product of the agents' state spaces \cite{puterman2014markov}, whose memory and running time grow exponentially with the number of agents. The key research gap we will address in this paper is to certify a lower bound for multi-agent systems under such specifications without this exponential cost.

\par Existing certified methods cope with this cost in different ways. Control barrier certificates certify the continuous
system directly, but the certificate is a function of the joint
state, and its search grows quickly with the dimension \cite{jagtap2020formal}. Despite recent advances, they also remain limited mostly to simple specifications, such as safety and reach-avoid. Among abstraction-based methods, orthogonally decoupled interval MDPs (odIMDPs) \cite{mathiesen2025scalable} store the transition bounds of each state-input pair per dimension. This reduces memory by orders of magnitude compared to standard interval MDP abstractions and handles coupled dynamics. However, odIMDP still computes the value function over the joint state space and maximizes over the joint input space.
A subset of the authors' previous work, SimRel tree \cite{wang2026correct}, abstracts each agent as an MDP and stores the value function as a tensor tree, whose memory grows linearly with the number of agents. It quantifies the abstraction error by an approximate simulation relation \cite{haesaert2020robust} as one probability that is subtracted from the lower bound at every step, regardless of the state. This uniform deduction makes the lower bound conservative.
\par \noindent {\bfseries Contributions.} This paper presents RMDP tree, a method for controller synthesis of multi-agent RMDPs with decoupled dynamics under temporal logic specifications. RMDP tree combines the advantages of odIMDPs and SimRel tree. As in odIMDPs, the uncertainty is modelled by ambiguity sets of transition distributions, and it lowers the value only through the transitions it affects. This yields tighter lower bounds than SimRel tree, which charges the same abstraction error in every state. As in SimRel tree, the value function is stored as a tensor tree, which avoids the joint state and input spaces of odIMDPs. First, we model each agent by its own ambiguity sets, and show that robust dynamic programming over the transitions of the specification automaton yields a lower bound on the satisfaction probability that holds for every choice of distributions from these sets. 
Second, we show that for decoupled agents, a lower bound on the satisfaction probability can be computed efficiently as a canonical polyadic decomposition (CPD), stored in the tree. 
Our case studies confirm both effects: on a four-agent benchmark, RMDP tree needs orders of magnitude less memory than odIMDP, and it certifies higher lower bounds than SimRel tree at the same resolution.

\section{Preliminaries and problem setup}
\subsection{Notations}
For a set $\mathbb S$, we denote its cardinality by $|\mathbb S|$. 
For a finite set $\spaceS$, we denote the set of probability distributions on $\spaceS$ by $\mathcal P(\spaceS)$, and write $s\sim \gamma$ for an element $s\in \spaceS$ drawn from $\gamma\in \mathcal P(\spaceS)$. For vectors $\mathrm{v}_i\in \mathbb{R}^{n_i}$ we define $\operatorname{col}(\mathrm{v}_1,\ldots,\mathrm{v}_N):=[\mathrm{v}_1^{\top}\ldots \mathrm{v}_N^{\top}]^\top \in\mathbb{R}^{\sum_{i=1}^N n_i}$, and their outer product $\mathrm v_1\otimes\dots\otimes \mathrm v_N\in\mathbb R^{n_1\times\dots\times n_N}$ as the order-$N$ array whose entry
at index $(j_1,\dots,j_N)$, with $j_i\in[1,n_i]$, is $\prod_{i=1}^N \mathrm v_i(j_i)$, where $\mathrm v_i(j_i)$ denotes the $j_i$-th
entry of $\mathrm v_i$. The indicator function is $\indicator_A(a):=1$ if $a\in A$, and $0$ otherwise. 

\subsection{Models: stochastic multi-agent systems}
We consider multi-agent systems of $N$ dynamically decoupled agents: the state evolution of each agent depends only on its own state and its own input. 
In this paper, we model each agent as a robust Markov decision process: 
\begin{definition}
    [robust Markov decision processes]\label{def:RMDP}
\hspace{1mm}A robust Markov decision process (RMDP) is a tuple $\mdpM=(\spaceS,\spaceU,\Gamma,\spaceY,s_0,g)$ with
\begin{itemize}
    \item a finite set of states $\spaceS$ with states $s\in\spaceS$;
    \item a finite set of inputs $\spaceU$ with inputs $u\in\spaceU$;
    \item for every $(s,u)\in \spaceS \times \spaceU$ a nonempty compact set $\Gamma(s,u)\subseteq \mathcal P(\spaceS)$ of transition distributions, the so-called ambiguity set of $(s,u)$;
    \item an output space $\spaceY$ with outputs $y\in\spaceY$;
    \item an initial state $s_0\in\spaceS$;
    \item an output map $g:\spaceS \rightarrow \spaceY$.
\end{itemize}
\end{definition}
The ambiguity set $\Gamma(s,u)$ collects the transition distributions the model admits at $(s,u)$. It captures the error of a finite-state abstraction of a continuous-state system, as well as parametric or distributional uncertainty in the model. If $\Gamma(s,u)$ is a set of probability vectors between a lower and an upper bound, the RMDP is an interval MDP \cite{nilim2005robust,givan2000bounded,delimpaltadakis2023interval}, also referred to as bounded-parameter MDP. Interval MDPs arise, for instance, as abstractions of continuous-state systems \cite{lahijanian2015formal}. If every $\Gamma(s,u)$ is a singleton, that is, the transition distribution is known exactly, the RMDP reduces to a finite MDP in the sense of \cite{haesaert2016verification}. 

\par The $i$-th agent is modeled as an RMDP
\begin{align}
    \mdpM_i=(\spaceS_i,\spaceU_i,\Gamma_i,\spaceY_i,s_{i,0},g_i ), \quad  i \in [ 1,N ], \label{eq:agent_RMDP}
\end{align}
and the multi-agent system is obtained through Cartesian composition of the agents:
\begin{align} \label{eq:cart_RMDP}
    \mdpM=(\spaceS,\spaceU,\Gamma,\spaceY,s_0,g )
\end{align}
with $\spaceS=\prod_{i=1}^{N}\spaceS_i$, $\spaceU=\prod_{i=1}^{N}\spaceU_i$,$\Gamma(s,u):=\{ \gamma\in \mathcal P(\spaceS) \mid \gamma(s')=\prod_{i=1}^N \gamma_i (s'_i) \text{ for all }s'\in \spaceS,\gamma_i\in \Gamma_i(s_i,u_i)   \}$, $\spaceY=\prod_{i=1}^{N}\spaceY_i$, $s_0:=\operatorname{col}(s_{1,0},\ldots,s_{N,0})$, and $g=\operatorname{col}(g_{1},\ldots,g_{N})$. 
The class of RMDPs is closed under this composition, and $\mdpM$ is itself an RMDP. 
\par The \textit{execution} of $\mdpM$ is a sequence of states $\boldsymbol{s}_{[0,T]}:=\{s[t] \mid t=0,\ldots,T\}$ initialized with $s[0]=s_0$. Consecutive states are realizations $s[t+1]\sim \gamma[t]$, with $\gamma[t]\in \Gamma(s[t],u[t])$. The \textit{output trajectory} is the sequence $\boldsymbol{y}_{[0,T]}:=\{y[t]\mid t=0,\ldots,T\}$ with $y[t]=g(s[t])$.
\par The input is selected by a policy and the distribution by an adversary. A deterministic policy $\boldsymbol{\pi}\in \Pi$ is a sequence $\boldsymbol{\pi}:=\{  \pi[t]|t=0,1,\ldots\}$ of maps that assign an input $u[t]=\pi[t](\history[t])\in \spaceU$ to each finite history $\history[t]:=s[0],u[0],\ldots,u[t-1],s[t]$. A deterministic adversary $\boldsymbol{\xi} \in \Xi$ is a sequence $\boldsymbol{\xi}:=\{  \xi[t]|t=0,1,\ldots\}$ of maps that assign a distribution $\gamma[t]=\xi[t](\history[t],u[t]) \in \Gamma(s[t],u[t])$ to each history and input. The adversary may depend on the whole history. In particular, it may choose a different distribution at every visit of the same state. Applying $\boldsymbol{\pi}\in\Pi$ and $\boldsymbol{\xi}\in\Xi$ to $\mdpM$ yields the controlled system $\mdpM \times \boldsymbol{\pi}\times\boldsymbol{\xi}$. For an initial state $s_0$, the controlled system defines a probability measure $\mathbb P_{\mdpM \times \boldsymbol{\pi}\times\boldsymbol{\xi}} ^{s_0}$ over its executions  \cite[Prop.~7.45]{bertsekas1996stochastic}.

\subsection{Specifications}
To each agent $\mdpM_i$, we associate a set of atomic propositions $\AP_i$, with $\AP_i\cap\AP_j=\emptyset$ for $i \neq j$. Let $\AP:=\bigcup_i \AP_i$ and $\Sigma:= 2^{\AP}$. The labeling map $L_i:\spaceY_i\rightarrow 2^{\AP_i}$ returns the propositions of agent $i$ that are true at the output $y_i$. The output trajectory $\boldsymbol{y}=y[0],y[1],\ldots$ of system \eqref{eq:cart_RMDP} is mapped to the word $\word:=l_0,l_1,\ldots$ by the labelling map $L:\spaceY \rightarrow \Sigma$, $L\bigl( \operatorname{col}(y_1 \dots y_N) \bigr):= \bigcup_{i=1}^N L_i(y_i)\in 2^{\AP}$, which translates each output to a letter $\letter_t=L(y[t])$.

\par
We consider temporal specifications $\phi$, e.g., syntactically co-safe linear temporal logic (scLTL) \cite{kupferman2001model} and linear temporal logic on finite traces (LTL\textsubscript{f}) \cite{de2013linear}. We write $\word \models \phi$ if the word $\word$ satisfies $\phi$ under the semantics of scLTL or LTL\textsubscript{f}.\ 
\par \noindent
\textbf{\textit{Problem statement:}}\hspace{1mm}\textit{
Given a multi-agent RMDP $\mdpM$ \eqref{eq:cart_RMDP}, a specification $\phi$, and a probability threshold $p\in[0,1]$, compute a policy $\boldsymbol{\pi}\in \Pi$ such that for all adversaries $\boldsymbol{\xi} \in \Xi$ }
    \begin{align}
    \ \mathbb{P}^{s_0}_{\!\mdpM \times \boldsymbol{\pi}\times\boldsymbol{\xi}}(\word\models \phi) \geq p.\label{eq:pb_statement}
\end{align}
A policy satisfying \eqref{eq:pb_statement} is certified against every history-dependent choice of transition distributions from the ambiguity sets. Dynamic programming over the joint state space solves this problem, but its memory and running time grow exponentially with the number of agents \cite{mathiesen2025scalable,abate2008probabilistic}. In this paper, we synthesize a policy and compute a lower bound on the probability that the system under this policy satisfies $\phi$, without constructing the joint state space. Whenever this lower bound is at least $p$, the policy satisfies \eqref{eq:pb_statement}. Sec.~\ref{sec:rdp} computes a lower bound by robust dynamic programming over the transitions of the DFA. Sec.~\ref{sec:tree} lower bounds it via a tensor tree that avoids the joint state space.

\section{Robust dynamic programming }
\label{sec:rdp}
\subsection{Specification satisfaction as a probabilistic reach-avoid problem}
scLTL and LTL\textsubscript{f} specifications translate into a deterministic finite automaton (DFA) via off-the-shelf tools \cite{duret2004spot}.
\begin{definition}[DFA]
\label{def:DFA} \hspace{1mm}
    A deterministic finite automaton (DFA) is defined by the tuple $\DFA=(Q,q_0,\Sigma,\tau_{\DFA},q_f)$ with a finite set of states $Q$, an initial state $q_0$, the alphabet $\Sigma$, a transition function $\tau_\DFA:(Q\setminus \{q_f\}) \times \Sigma \rightarrow Q$, and an accepting state $q_f$. A state $q_s\in Q\setminus\{q_f\}$ with $\tau_\DFA(q_s,\word)=q_s$ for all $\word\in \Sigma$ is a sink.
\end{definition}
We denote the DFA of a specification $\phi$ by $\DFA_\phi$ and drop the subscript in the sequel. For a word $\word=\letter_0,\letter_1,\ldots$, the run of $\DFA$ is the sequence $q_0,q_1,\ldots$ with $q_{t+1}=\tau_\DFA(q_t,\letter_t)$. The prefix $\word_{[0,T]}:=\letter_0,\ldots,\letter_T$ is accepted if the run reaches $q_f$ while reading it, that is, if $q_t=q_f$ for some $t\in [0,T+1]$, and we write $\word_{[0,T]} \models \phi$. A word satisfies $\phi$ if and only if a prefix of it is accepted \cite{kupferman2001model,de2013linear}. A run that reaches $q_f$ within $\word_{[0,T]}$ also reaches $q_f$ within $\word_{[0,T+1]}$. The set of words whose prefix $\word_{[0,T]}$ is accepted is therefore contained in the set of words whose prefix $\word_{[0,T+1]}$ is accepted. As $T$ grows without bound, these sets increase to the set of words that satisfy $\phi$ within an unbounded horizon, as in \eqref{eq:pb_statement}.

\par \noindent {\bfseries Transitions labelled by Boolean formula.} A Boolean formula $\subphi$ is a conjunction of possibly negated atomic propositions. A letter $\letter$ satisfies $\subphi$, written as $\letter \models \subphi$, if $\letter$ contains every proposition that $\subphi$ requires to be true and no proposition that $\subphi$ requires to be false. For $q,q'\in Q$, we label the transition from $q$ to $q'$ with a set of Boolean formulas such that each letter $\letter$ with $\tau_\DFA(q,l)=q'$ satisfies exactly one Boolean formula of the set, and each letter $\letter$ with $\tau_\DFA(q,l)\neq q'$ satisfies none. We write $\tau_\DFA(q,\subphi):=q'$ for every Boolean formula $\subphi$ in this set.   
Since the agents' propositions are disjoint, every Boolean formula splits into one Boolean formula per agent: $\subphi=\subphi_1 \wedge \ldots \wedge \subphi_N$, where $\subphi_i$ collects the literals of $\subphi$ over $\AP_i$ ($\subphi_i =\top$ if there are none). 

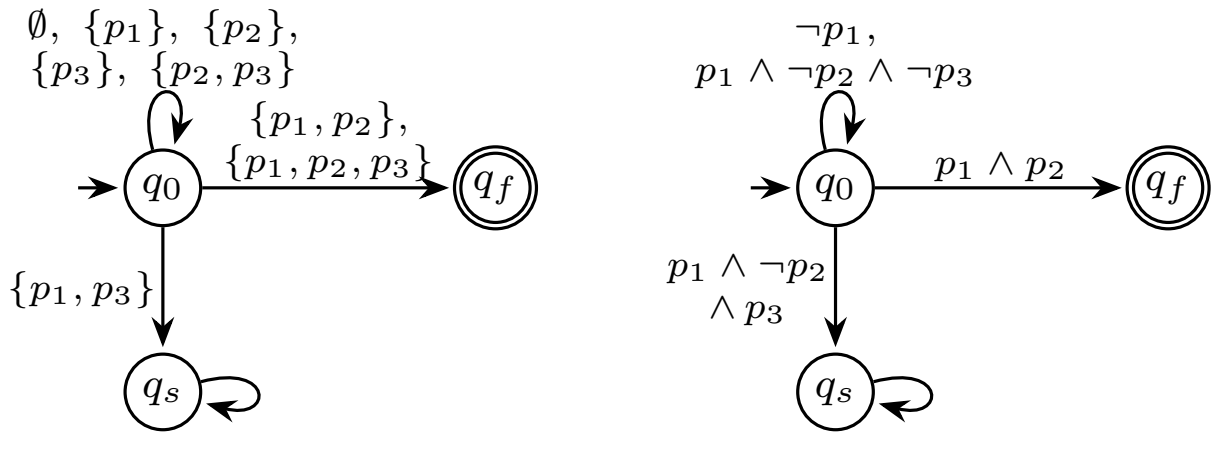
\begin{figure}[htbp]
\centering
\tikzset{dfa/.style={>=Stealth, shorten >=1pt, semithick, auto,
    every state/.style={minimum size=5mm, inner sep=0pt, font=\footnotesize},
    initial text={}, initial distance=3mm},
  lab/.style={font=\scriptsize, align=center, inner sep=1pt}}
\begin{subfigure}[b]{0.49\columnwidth}\centering
\begin{tikzpicture}[dfa]
\node[state, initial] (q0) at (0,0) {$q_0$};
\node[state, accepting] (qf) at (2.2,0) {$q_f$};
\node[state] (qs) at (0,-1.35) {$q_s$};
\path[->] (q0) edge[loop above] node[lab] {$\emptyset,\ \{p_1\},\ \{p_2\},$\\$\{p_3\},\ \{p_2,p_3\}$} (q0);
\path[->] (q0) edge node[lab] {$\{p_1,p_2\},$\\$\{p_1,p_2,p_3\}$} (qf);
\path[->] (q0) edge node[lab,swap] {$\{p_1,p_3\}$} (qs);
\path[->] (qs) edge[loop right]  (qs);
\end{tikzpicture}
\caption{Letters}
\label{subfig:dfa_letter}
\end{subfigure}\hfill
\begin{subfigure}[b]{0.49\columnwidth}\centering
\begin{tikzpicture}[dfa]
\node[state, initial] (q0) at (0,0) {$q_0$};
\node[state, accepting] (qf) at (2.2,0) {$q_f$};
\node[state] (qs) at (0,-1.35) {$q_s$};
\path[->] (q0) edge[loop above] node[lab] {$\neg p_1,$\\$p_1\wedge\neg p_2\wedge\neg p_3$} (q0);
\path[->] (q0) edge node[lab] {$p_1\wedge p_2$} (qf);
\path[->] (q0) edge node[lab,swap] {$p_1\wedge\neg p_2$\\$\wedge\, p_3$} (qs);
\path[->] (qs) edge[loop right] (qs);
\end{tikzpicture}
\caption{Boolean formulas}
\label{subfig:dfa_subformula}
\end{subfigure}
\caption{DFA of Example~\ref{example} with transitions labelled by (a) letters and (b) Boolean formulas. $q_s$ is the sink.}
\label{fig:ex_dfa}
\end{figure}

\begin{example}\label{example}\hspace{1mm}
    Consider two agents with $y_1\in \spaceY_1=[-10,10]$ the output of agent 1, and $y_2\in\spaceY_2= [-10,10]$ the output of agent 2, respectively, and propositions $\AP_1=\{p_1\}$ and $\AP_2=\{p_2,p_3\}$, where $p_1$ is true iff $y_1\in [4,10]$, $p_2$ iff $y_2\in[-4,0]$, and $p_3$ iff $y_2\in [0,4]$. The specification $\phi:=\neg (p_1 \wedge p_3) \until (p_1\wedge p_2)$ asks the multi-agent system to reach $p_1\wedge p_2$ while avoiding $p_1 \wedge p_3$. Fig.~\ref{fig:ex_dfa} shows its DFA. The eight letters leaving $q_0$ in Fig.~\ref{subfig:dfa_letter} reduce to four Boolean formulas in Fig.~\ref{subfig:dfa_subformula}. The Boolean formula $\neg p_1$ splits into $\subphi_1 =\neg p_1$ and $\subphi_2=\top$. The output trajectory $\boldsymbol{y}=[3.5,0.5]^\top,[4.2,-0.5]^\top$ generates the word $\{p_3\},\{p_1,p_2\}$. Starting at $q_0$, the letter $\{p_3\}$ satisfies $\neg p_1$, so the run stays in $q_0$. The next letter $\{p_1,p_2\}$ satisfies $p_1\wedge p_2$, so the run moves to $q_f$, and the word is accepted.
\end{example}
An execution of $\mdpM \times \boldsymbol{\pi} \times\boldsymbol{\xi}$ satisfies $\phi$ if and only if the run of $\DFA$ on its word $\word$ reaches $q_f$. Hence, the probability of satisfying $\phi$ is the probability of reaching $q_f$ on the product of $\mdpM$ and $\DFA$ under $\boldsymbol{\pi}$ and $\boldsymbol{\xi}$ \cite{tkachev2017quantitative,haesaert2020robust}.

\subsection{Dynamic programming}
The DFA state summarizes the output history that is relevant to the satisfaction of $\phi$. As in \cite{wang2025unraveling}, we call it the \emph{mode} and use it as an index: the policy and the adversary are families with one member per mode. A mode-dependent policy $\boldsymbol{\pi}:=\{ \boldsymbol{\pi}_q \mid q\in Q\}$ has, for every mode $q$, a sequence $\boldsymbol{\pi}_q:=\{ \pi_q[t] | t=0,1,\ldots\}$ of maps $\pi_q[t]:\spaceS\rightarrow \spaceU$. A mode-dependent adversary $\boldsymbol{\xi}:=\{ \boldsymbol{\xi}_q \mid q\in Q\}$ has, for every mode $q$, a sequence $ \boldsymbol{\xi}_q:=\{\xi_q[t]|t=0,1,\ldots \}$ of maps $\xi_q[t]:\spaceS \times \spaceU \rightarrow \mathcal P(\spaceS)$ with $\xi_q[t](s,u)\in \Gamma(s,u)$ for all $(s,u) \in \spaceS \times \spaceU$. The mode-dependent policy follows \cite{wang2026correct}. The mode-dependent adversary follows \cite{mathiesen2025scalable}, with the input, the mode and the time index as additional arguments. During execution, the run of $\DFA$ reaches $q_{t+1}$ after reading $\letter_0,\ldots,\letter_t$. The policy chooses the input $u[t]=\pi_{q_{t+1}}[t](s[t])$, and the adversary chooses the distribution $\gamma[t]=\xi_{q_{t+1}}[t](s[t],u[t])$. Since $q_{t+1}$ is determined by $s[0],\ldots,s[t]$, $\boldsymbol{\pi}\in \Pi$ and $\boldsymbol{\xi} \in \Xi$. 
\par On the product $\mdpM \times \DFA$ \cite{baier2008principles} with product states $(s,q)\in \spaceS \times Q$, mode-dependent $\boldsymbol{\pi}$ and $\boldsymbol{\xi}$ are Markov, and a prefix of the word is accepted if and only if the product reaches $\spaceS\times \{q_f\}$ \cite{kupferman2001model,baier2008principles}. For a mode-dependent policy $\boldsymbol{\pi}$ and adversary $\boldsymbol{\xi}$, we define the value function $\tensorV_{q,t}^{\boldsymbol{\pi},\boldsymbol{\xi}}(s)$ as the probability of reaching $\spaceS\times \{q_f\}$ within $t$ steps from $(s,q)$ \cite{abate2008probabilistic}:
$$
\tensorV_{q,t}^{\boldsymbol{\pi},\boldsymbol{\xi}}(s):=\mathbb P_{(\mdpM \times \DFA) \times \boldsymbol{\pi} \times \boldsymbol{\xi}}^{(s,q)} (\exists k \leq t: \bar{s}[k] \in \spaceS \times \{q_f\})
$$
with $\bar{s}[k]$ the product state $k$ steps after $(s,q)$. Here the product is in $(s,q)$ at time $T-t$, so the step from $\bar{s}[k]$ to $\bar{s}[k+1]$ follows $\pi[T-t+k]$ and $\xi[T-t+k]$. Initialized as $\tensorV_{q,0}=1$ if $q=q_f$, and $0$ otherwise, the value functions are computed iteratively as \cite{puterman2014markov}
\begin{equation}\label{eq:vi_fixed_pol_adv}
	\begin{aligned}
		\!\!&\tensorV_{q,t+1}^{\boldsymbol{\pi},\boldsymbol{\xi}}(s)= \indicator_{q_f}(q)
 +\indicator_{Q\setminus\{q_f\}}(q)
		\expectation_{s'\sim \gamma}[\tensorV_{q',t}^{\boldsymbol{\pi},\boldsymbol{\xi}}(s')], \!\!
	\end{aligned}
\end{equation} where $\gamma:=\xi_q[T-1-t](s,u)$ with $u=\pi_q[T-1-t](s)$ is the distribution of the next state $s'$, $q':=\tau_\DFA(q,L(g(s')))$ is the next DFA state, and $\expectation_{s'\sim \gamma}[\tensorV_{q',t}^{\boldsymbol{\pi},\boldsymbol{\xi}}(s')]=\sum_{s'\in\spaceS} \gamma(s')\tensorV_{q',t}^{\boldsymbol{\pi},\boldsymbol{\xi}}(s')$.

\begin{proposition}[Specification satisfaction \cite{haesaert2020robust}] \label{prop:spec_satisfaction}
\hspace{1mm}For a mode-dependent policy $\boldsymbol{\pi}$ and a mode-dependent adversary $\boldsymbol{\xi}$ of horizon $T$, that is, with $t=0,1,\ldots,T-1$
\begin{equation} \label{eq:prop1}
\mathbb{P}^{s_0}_{\mdpM \times \boldsymbol{\pi}\times\boldsymbol{\xi}}(\word_{[0,T]} \models \phi )=\tensorV_{\bar{q}_0,T}^{\boldsymbol{\pi},\boldsymbol{\xi}}(s_0)
\end{equation} with $\bar q_0= \tau_{\mathcal A}(q_0,L(g(s_0)))$, and $\tensorV^{\boldsymbol{\pi},\boldsymbol{\xi}}$ computed as in \eqref{eq:vi_fixed_pol_adv}.
\end{proposition}
Since accepted prefixes stay accepted, the probability in \eqref{eq:prop1} is nondecreasing in $T$ and as $T\rightarrow \infty$, converges to the specification satisfaction probability $\mathbb P_{\mdpM \times \boldsymbol{\pi} \times \boldsymbol{\xi}}^{s_0}(\word \models \phi) $ in \eqref{eq:pb_statement} \cite{tkachev2017quantitative}. 
\par
The expectation in \eqref{eq:vi_fixed_pol_adv} sums over the next states $s'$, each leading to its own $q'$ starting from $q$. We could equivalently sum over the transitions leaving $q$, collected with their target $q'$ in $\text{Succ}(q):=\{  (\subphi,q') | q'=\tau_\DFA(q,\subphi)\}$. For instance in Example~\ref{example}, $\text{Succ}(q_0)=\{ (\neg p_1,q_0), (p_1\wedge \neg p_2 \wedge \neg p_3, q_0), (p_1\wedge p_2, q_f), (p_1\wedge \neg p_2 \wedge p_3,q_s) \}$, so the sum starting from $q_0$ has four items.  
Since the Boolean formulas leaving $q$ partition $\Sigma$, the letter $L(g(s'))$ of every next state $s'$ satisfies exactly one of them, and $s'$ leads to its target DFA state. For instance in Example~\ref{example}, the letter of $y'=[4.2,-0.5]^\top$ satisfies $p_1\wedge p_2$ and leaves from $q_0$ to $q_f$. We extend the DFA-informed operator of \cite{wang2025unraveling} with the adversary:
\begin{align}\label{eq:op_general}
\op_{\subphi}^{\pi_q,\xi_q}(\tensorV)(s):=\sum_{s'\in\spaceS } \indicator[L(g(s')) \models \subphi] \gamma(s') \tensorV(s'),
\end{align}
where $\indicator[L(g(s')) \models \subphi]:=1$ if $L(g(s')) \models \subphi$, and $0$ otherwise, and $\gamma=\xi_q(s,\pi_q(s))$ is the distribution that the adversary chooses at $s$ for the input $\pi_q(s)$. The DFA-informed operator thus lets the recursion sum over the transitions leaving $q$ instead of over the next states $s'$. Each transition collects the next states whose letter satisfies its formula. Lemma~\ref{lemma:dfavi_equi} states that this recursion equals \eqref{eq:vi_fixed_pol_adv}. 

\begin{lemma} \label{lemma:dfavi_equi}
    \hspace{1mm}For $q\in Q\setminus \{q_f\}$, the recursion in \eqref{eq:vi_fixed_pol_adv} is equivalent to 
    \begin{tightdisplay}
    \begin{align} \label{eq:Viterations}
\tensorV_{q,t+1}^{\boldsymbol{\pi},\boldsymbol{\xi}}(s)=\sum_{(\subphi,q')\in {\small \text{Succ}}(q)}\op_{\subphi}^{\pi_q[T-1-t],\xi_q[T-1-t]}(\tensorV^{\boldsymbol{\pi},\boldsymbol{\xi}}_{q',t})(s)
\end{align}
\end{tightdisplay}
with the same initialization as \eqref{eq:vi_fixed_pol_adv} and $\tensorV^{\boldsymbol{\pi},\boldsymbol{\xi}}_{q_f,t}=1$ for all $t$.
\end{lemma}
\begin{proof}
    The proof follows that of \cite[Lemma~1]{wang2025unraveling} with the transition kernel of the MDP replaced by the distribution that the adversary chooses.
\end{proof}

 
\subsection{Robust dynamic programming}
We define \emph{robust value functions} $\underline{\tensorV}_{q,t}^{\boldsymbol{\pi}}:\spaceS \rightarrow [0,1]$ by letting the adversary minimize at every step of \eqref{eq:Viterations}, starting from $\underline{\tensorV}_{q,0}=0$ for $q\in Q \setminus \{q_f\}$ and $\underline{\tensorV}_{q_f,t}=1$:
\begin{equation} \label{eq:robust_valuefunc}
    \begin{aligned}
        \underline{\tensorV}_{q,t+1}^{\boldsymbol{\pi}}(s)
         =\min_{\xi_q[T-1-t]} \!\!\!\!\!\!\!\sum_{(\subphi,q')\in \text{Succ}(q)} \!\!\!\!\!\!\!\!\!\!\op_{\subphi}^{\pi_q[T-1-t],\xi_q[T-1-t]}(\underline{\tensorV}_{q',t}^{\boldsymbol{\pi}})(s),
    \end{aligned}
\end{equation}
where the minimum is over $\xi_q[T-1-t](s,u)\in \Gamma(s,u)$ with $u=\pi_q[T-1-t](s)$.
\begin{theorem}[Robust value function as a lower bound]\label{theorem:robust_lb}
\hspace{1mm}Given a mode-dependent policy $\boldsymbol{\pi}$ of horizon $T$, that is, with $t=0,1,\ldots,T-1$, the probability that the prefix $\word_{[0,T]}$ is accepted is lower bounded for any $\boldsymbol{\xi}\in\Xi$ as
\begin{equation} \label{eq:theorem1_lb}
     \mathbb P_{\mdpM \times \boldsymbol{\pi}\times\boldsymbol{\xi}} ^{s_0} (\word_{[0,T]} \models \phi) \geq \underline{\tensorV}_{\bar{q}_0,T}^{\boldsymbol{\pi}} (s_0),
\end{equation}
with $\bar{q}_0=\tau_\DFA(q_0,L(g(s_0)))$ and $\underline{\tensorV}^{\boldsymbol{\pi}}$ computed as in \eqref{eq:robust_valuefunc}.
\end{theorem}
Thus, the robust value function $\underline{\tensorV}_{\bar{q}_0,T}^{\boldsymbol{\pi}} (s_0)$ serves as a lower bound on the specification satisfaction probability $\mathbb P_{\mdpM \times \boldsymbol{\pi}\times\boldsymbol{\xi}} ^{s_0} (\word_{[0,T]} \models \phi)$, under any possible realization $\boldsymbol{\xi}$ of the uncertainty. In fact, since $\mathbb P_{\mdpM \times \boldsymbol{\pi} \times \boldsymbol{\xi}} ^{s_0}(\word \models \phi) \geq \mathbb \mathbb P_{\mdpM \times \boldsymbol{\pi} \times \boldsymbol{\xi}} ^{s_0}(\word_{[0,T]} \models \phi)$ for every $T$, the lower bound in \eqref{eq:theorem1_lb} also lower bounds the specification satisfaction probability $\mathbb P_{\mdpM \times \boldsymbol{\pi} \times \boldsymbol{\xi}} ^{s_0}(\word \models \phi)$ in \eqref{eq:pb_statement}, and $\boldsymbol{\pi}$ solves the problem as long as $\underline{\tensorV}_{\bar{q}_0,T}^{\boldsymbol{\pi}}(s_0) \geq p$.

\section{Robust tensor tree}\label{sec:tree}
Computing $\underline{\tensorV}$ in \eqref{eq:robust_valuefunc} is computationally heavy. For $\spaceS =\prod_{i=1}^N \spaceS_i$, each value function $\underline{\tensorV}_{q,t}:\spaceS \rightarrow[0,1]$ is an order-$N$ tensor, that is, an array in $\spaceR^{|\spaceS_1| \times \ldots \times |\spaceS_N|}$. Storing it entrywise requires $\prod_{i=1}^N |\spaceS_i|$ entries. For instance, it would be $10^{24}$ entries for $8$ agents with $10^3$ states per agent. Each entry of \eqref{eq:robust_valuefunc} further requires a minimization over the joint ambiguity set $\Gamma(s,u)$ of a sum over all $\prod_{i=1}^N |\spaceS_i|$ successor states $s'$. Memory and running time therefore grow exponentially in the number of agents. A tensor with low-rank structure, however, can be stored far more compactly.
\begin{figure}[htbp]
\centering
\resizebox{0.8\columnwidth}{!}{%
 
\begin{tikzpicture}[line join=round]
\tikzset{cell/.style={draw=black,minimum size=2.4mm,inner sep=0pt}}
\newcommand{\startwo}[3]{%
 \begin{scope}[shift={(#1,#2)}]
 \foreach \dx/\dy/\col/\d in {0/1/blue/1, 1/0/red/2}{
   \foreach \k in {1,...,4}{\pgfmathtruncatemacro{\v}{20+mod(\k*37+#3*53+\d*29,66)}
     \node[cell,fill=\col!\v] at ({\k*0.25*\dx},{\k*0.25*\dy}) {};}
   }
 \node[cell,fill=black!20] at (0,0) {};
 \end{scope}}
\newcommand{\starthree}[3]{%
 \begin{scope}[shift={(#1,#2)}]
 \foreach \dx/\dy/\col/\d in {0/1/blue/1, 1/0/red/2, -0.75/-0.6/green!60!black/3}{
   \foreach \k in {1,...,4}{\pgfmathtruncatemacro{\v}{20+mod(\k*37+#3*53+\d*29,66)}
     \node[cell,fill=\col!\v] at ({\k*0.25*\dx},{\k*0.25*\dy}) {};}
   }
 \node[cell,fill=black!20] at (0,0) {};
 \end{scope}}
\newcommand{\starfour}[3]{%
 \begin{scope}[shift={(#1,#2)}]
 \foreach \dx/\dy/\col/\d in {0/1/blue/1, 1/0/red/2, -0.75/-0.6/green!60!black/3, -0.75/0.6/orange/4}{
   \foreach \k in {1,...,4}{\pgfmathtruncatemacro{\v}{20+mod(\k*37+#3*53+\d*29,66)}
     \node[cell,fill=\col!\v] at ({\k*0.25*\dx},{\k*0.25*\dy}) {};}
   }
 \node[cell,fill=black!20] at (0,0) {};
 \end{scope}}
\begin{scope}[shift={(0,1.8)}]
 \begin{scope}[shift={(0,0.08)},scale=1.24]
 \def\s{0.25}
 \foreach \i in {0,...,3}{\foreach \j in {0,...,3}{
   \pgfmathsetmacro{\c}{20+12*\i+10*\j}
   \fill[green!\c!blue] (0.15+\i*\s,\j*\s) rectangle ++(\s,\s); \draw[black!50,very thin] (0.15+\i*\s,\j*\s) rectangle ++(\s,\s);}}
 \draw (0.15,0) rectangle ++(1,1);
 \end{scope}
 \node at (2.05,0.6) {$=$};
 \startwo{2.75}{0.2}{1}
 \node at (4.13,0.6) {$+$};
 \startwo{4.50}{0.2}{2}
 \node at (5.88,0.6) {$+$};
 \startwo{6.25}{0.2}{3}
 \node at (7.63,0.6) {$+$};
 \startwo{8.00}{0.2}{4}
\end{scope}
\begin{scope}[shift={(0,0)}]
 \begin{scope}[shift={(0,-0.27)},scale=1.36]
 \def\s{0.25}
 \foreach \i in {0,...,3}{\foreach \j in {0,...,3}{
   \pgfmathsetmacro{\c}{20+12*\i+10*\j}
   \fill[green!\c!blue] (\i*\s,\j*\s) rectangle ++(\s,\s); \draw[black!50,very thin] (\i*\s,\j*\s) rectangle ++(\s,\s);
   \fill[green!\c!red!80!black] ({1+\i*0.09},{\j*\s+\i*0.07}) -- ++(0.09,0.07) -- ++(0,\s) -- ++(-0.09,-0.07) -- cycle;
   \draw[black!50,very thin] ({1+\i*0.09},{\j*\s+\i*0.07}) -- ++(0.09,0.07) -- ++(0,\s) -- ++(-0.09,-0.07) -- cycle;
   \fill[green!\c!yellow] ({\j*\s+\i*0.09},{1+\i*0.07}) -- ++(\s,0) -- ++(0.09,0.07) -- ++(-\s,0) -- cycle;
   \draw[black!50,very thin] ({\j*\s+\i*0.09},{1+\i*0.07}) -- ++(\s,0) -- ++(0.09,0.07) -- ++(-\s,0) -- cycle;}}
 \end{scope}
 \node at (2.05,0.6) {$=$};
 \starthree{3.25}{0.45}{1}
\end{scope}
\begin{scope}[shift={(0,-2.15)}]
 \def\a{0.62}\def\b{0.28}
 \begin{scope}[shift={(0.76,0.54)},scale=1.39]
 \def\a{0.55}\def\b{0.25}\def\ox{0.26}\def\oy{0.22}\def\ix{0.12}\def\iy{0.10}
 \coordinate (F1) at (-\a,-\a); \coordinate (F2) at (\a,-\a); \coordinate (F3) at (\a,\a); \coordinate (F4) at (-\a,\a);
 \coordinate (B1) at (-\a+\ox,-\a+\oy); \coordinate (B2) at (\a+\ox,-\a+\oy); \coordinate (B3) at (\a+\ox,\a+\oy); \coordinate (B4) at (-\a+\ox,\a+\oy);
 \coordinate (f1) at (-\b,-\b); \coordinate (f2) at (\b,-\b); \coordinate (f3) at (\b,\b); \coordinate (f4) at (-\b,\b);
 \coordinate (b1) at (-\b+\ix,-\b+\iy); \coordinate (b2) at (\b+\ix,-\b+\iy); \coordinate (b3) at (\b+\ix,\b+\iy); \coordinate (b4) at (-\b+\ix,\b+\iy);
 \fill[blue!70,opacity=0.45] (F1)--(F2)--(f2)--(f1)--cycle;
 \fill[green!60!black,opacity=0.40] (F2)--(F3)--(f3)--(f2)--cycle;
 \fill[yellow!80!green,opacity=0.50] (F3)--(F4)--(f4)--(f3)--cycle;
 \fill[teal,opacity=0.45] (F4)--(F1)--(f1)--(f4)--cycle;
 \fill[red!70!green,opacity=0.35] (F2)--(B2)--(B3)--(F3)--cycle;
 \fill[yellow!70!green,opacity=0.35] (F4)--(F3)--(B3)--(B4)--cycle;
 \fill[blue!60!green,opacity=0.8] (f1) rectangle (f3);
 \fill[green!50!black,opacity=0.7] (f2)--(b2)--(b3)--(f3)--cycle;
 \fill[yellow!70!green,opacity=0.8] (f4)--(f3)--(b3)--(b4)--cycle;
 \draw[thick] (F1)--(F2)--(F3)--(F4)--cycle; \draw[thick] (F2)--(B2)--(B3)--(B4)--(F4); \draw[thick] (F3)--(B3);
 \draw[thick,black!40] (F1)--(B1)--(B2); \draw[thick,black!40] (B1)--(B4);
 \draw[black!70] (f1)--(f2)--(f3)--(f4)--cycle; \draw[black!70] (f2)--(b2)--(b3)--(b4)--(f4); \draw[black!70] (f3)--(b3);
 \foreach \k in {1,2,3,4}{\draw[black!60,thin] (F\k)--(f\k);}
\end{scope}
 \node at (2.05,0.6) {$=$};
 \starfour{3.25}{0.5}{1}  \node at (4.55,0.6) {$+$};
 \starfour{5.65}{0.5}{2}
\end{scope}
\end{tikzpicture}
}
\caption{Canonical polyadic decomposition (CPD) of, from top to bottom, a rank-4 order-2 tensor, a rank-1 order-3 tensor and a rank-2 order-4 tensor (left), as a sum of rank-1 tensors (right). Each rank-1 tensor is the outer product of vectors, drawn as its arms.}
\label{fig:cpd}
\end{figure}
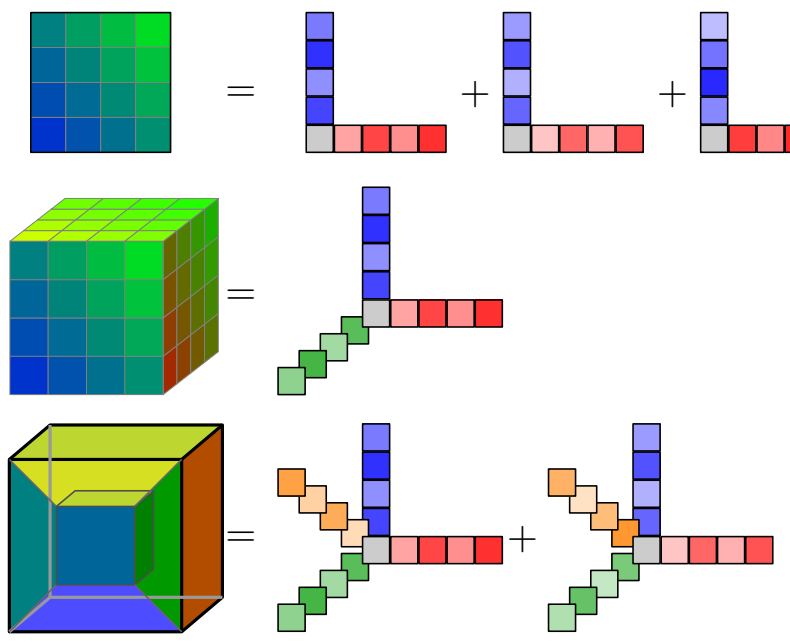
\par If an order-$N$ tensor $\tensorV\in \mathbb R^{|\spaceS_1| \times \ldots \times |\spaceS_N|}$ can be written as 
\begin{tightdisplay}
$$
\tensorV=\sum_{r=1}^R \mathrm v_r^1 \otimes \ldots \otimes \mathrm v_r^N,\quad \mathrm v_t^i\in \mathbb R^{|\spaceS_i|},
$$
\end{tightdisplay}
where $\otimes$ is the outer product, that is, $(\mathrm v^1_r \otimes \ldots \otimes \mathrm v^N_r)(s)=\prod_{i=1}^N \mathrm v^i_r(s_i)$ for $s=(s_1,\ldots,s_N)\in \mathbb S$, and $\mathrm v^i_r(s_i)$ is the entry of $\mathrm v_r^i$ at state $s_i$, then this sum is a rank-$R$ canonical polyadic decomposition (CPD) of $\tensorV$ \cite{hitchcock1927expression}. Fig.~\ref{fig:cpd} shows three CPDs. Storing $\tensorV$ in its CPD requires $O(R\sum_{i=1}^N |\spaceS_i|)$ memory instead of $O(\prod_{i=1}^N |\spaceS_i|)$. For instance, $O(R\times 8\times 10^3)$ instead of $O(10^{24})$. This is the idea of our approach: for decoupled agents, we compute a lower bound on $\underline{\tensorV}$ that is a CPD.

\subsection{RMDP tensor tree}
A prefix is accepted when its letters lead the run over the DFA to $q_f$ from $q_0$. Read backwards, each such run starts at $q_f$ and steps against the direction of the DFA transitions until it reaches $q_0$. Self-loops and cycles in the DFA give many such runs, of different lengths. 
\begin{definition}[Tensor tree $\mathcal{G}$ \cite{wang2025unraveling}] \label{def:r1tree}
\hspace{1mm}Given a DFA  $\DFA=(Q,q_0,\Sigma,\tau_{\DFA},q_f)$, 	a tensor tree
	$\mathcal{G}=(\mathcal{Z},\mathcal{E},\qmapping, \valuemapping)$ is a rooted tree, that is, a connected directed graph without cycles in which the root vertex has no parent and every other vertex has exactly one parent. It consists of \begin{itemize}
			\item a set of vertices $\mathcal{Z}$ with vertices $z\in \mathcal{Z}$;
			\item a set of directed edges $\mathcal{E}$ with edges $ (z,\subphi,z')\in\mathcal{E} $ from a parent $z$ to a child $z'$, labelled with a Boolean formula $\subphi$;
			\item a DFA-mode mapping $\qmapping:\mathcal{Z}\rightarrow Q$ that maps a vertex $z\in\mathcal{Z}$ to a DFA mode $q\in Q$, such that $\tau_\DFA(\mathcal L_Q(z'),\subphi)=\mathcal L_Q(z)$ for every edge $(z,\subphi,z')$;
			\item a value mapping $\valuemapping$ that maps a vertex $z\in \mathcal{Z}$ to a tensor $\valuemapping(z)\in \mathbb R^{|\spaceS_1|\times \ldots \times |\spaceS_N| }$.
	\end{itemize}
    A vertex without children is a leaf.
\end{definition}
\begin{figure}[htbp]
\centering
\definecolor{cN}{RGB}{200,228,195}  
\definecolor{cG}{RGB}{200,228,195}  
\definecolor{cY}{RGB}{255,252,190}  
\newcommand{\vstar}[3]{
 \begin{scope}[shift={(#1-0.25,#2-0.25)}]
 \foreach \dx/\dy/\col/\d in {0/1/blue/1, 1/0/red/2}{
   \foreach \k in {1,...,4}{\pgfmathtruncatemacro{\v}{20+mod(\k*37+#3*53+\d*29,66)}
     \node[draw=black,line width=0.2pt,fill=\col!\v,minimum size=1.4mm,inner sep=0pt] at ({\k*0.14*\dx},{\k*0.14*\dy}) {};}}
 \node[draw=black,line width=0.2pt,fill=black!20,minimum size=1.4mm,inner sep=0pt] at (0,0) {};
 \end{scope}}
\newcommand{\tedge}[3]{
 \draw[line width=6pt,#3] #1 -- #2;
 \draw[->,>=Stealth,semithick] #1 -- #2;}
\begin{tikzpicture}[>=Stealth,semithick,
  every state/.style={minimum size=5mm,inner sep=0pt,font=\footnotesize},
  flab/.style={font=\scriptsize,inner sep=1.5pt}]
\node[state,accepting] (qf) at (1.3,1.45) {$q_f$};
\node[state,initial,initial text={},initial distance=2.5mm,initial where=right] (q0) at (1.3,-0.2) {$q_0$};
\node[state,draw=black!35,text=black!35] (qs) at (1.3,-1.85) {$q_s$};
\draw[->] (q0) -- (qf);
\draw[->,black!35] (q0) -- (qs);
\draw[->] (q0) to[out=160,in=200,looseness=7] (q0);
\draw[->,black!35] (qs) to[out=-20,in=20,looseness=7] (qs);
\node[flab,fill=cY] at (1.3,0.65) {$p_1\!\wedge p_2$};
\node[flab,fill=cN,anchor=south east] at (0.55,-0.1) {$\neg p_1$};
\node[flab,fill=cG,anchor=north east,align=center] at (0.55,-0.2) {$p_1\!\wedge\!\neg p_2$\\$\wedge\neg p_3$};
\node[flab,fill=white,text=black!35,align=center] at (1.3,-1.03) {$p_1\!\wedge\!\neg p_2\!\wedge p_3$};
\begin{scope}[xshift=5.05cm,yshift=1.45cm]
\def\up{0.47}\def\dn{0.37}
\coordinate (r) at (0,0);
\coordinate (a) at (0,-1.1);
\coordinate (b1) at (-1.15,-2.2); \coordinate (b2) at (1.15,-2.2);
\coordinate (c1) at (-1.9,-3.3); \coordinate (c2) at (-0.63,-3.3);
\coordinate (c3) at (0.63,-3.3); \coordinate (c4) at (1.9,-3.3);
\tedge{($(r)+(0,-\dn)$)}{($(a)+(0,\up)$)}{cY}
\tedge{($(a)+(-0.15,-\dn)$)}{($(b1)+(0.1,\up)$)}{cN}
\tedge{($(a)+(0.15,-\dn)$)}{($(b2)+(-0.1,\up)$)}{cG}
\tedge{($(b1)+(-0.15,-\dn)$)}{($(c1)+(0.05,\up)$)}{cN}
\tedge{($(b1)+(0.15,-\dn)$)}{($(c2)+(-0.05,\up)$)}{cG}
\tedge{($(b2)+(-0.15,-\dn)$)}{($(c3)+(0.05,\up)$)}{cN}
\tedge{($(b2)+(0.15,-\dn)$)}{($(c4)+(-0.05,\up)$)}{cG}
\vstar{0}{0}{1}
\vstar{0}{-1.1}{2}
\vstar{-1.15}{-2.2}{3}\vstar{1.15}{-2.2}{4}
\vstar{-1.9}{-3.3}{5}\vstar{-0.63}{-3.3}{6}\vstar{0.63}{-3.3}{7}\vstar{1.9}{-3.3}{8}
\begin{scope}[font=\scriptsize]
\node[anchor=east] at (-0.27,0.32) {root};
\node[anchor=east] at (-0.27,-0.78) {$z_1$};
\node[anchor=east] at (-1.42,-1.88) {$z_2$};
\node[anchor=east] at (0.88,-1.88) {$z_3$};
\node[anchor=east] at (-2.17,-2.98) {$z_4$};
\node[anchor=east] at (-0.90,-2.98) {$z_5$};
\node[anchor=east] at (0.36,-2.98) {$z_6$};
\node[anchor=east] at (1.63,-2.98) {$z_7$};
\end{scope}
\end{scope}
\end{tikzpicture}
\caption{Left: the DFA of Example~\ref{example}. Right: its tensor tree $\mathcal G_3$ without pruning, whose edge colors match the DFA transitions. The sink $q_s$ (grey) is excluded because no run from $q_s$ reaches $q_f$. Each vertex's value is the outer product of a vector of agent 1 (blue arm) and of agent 2 (red arm).}
\label{fig:tree}
\end{figure}
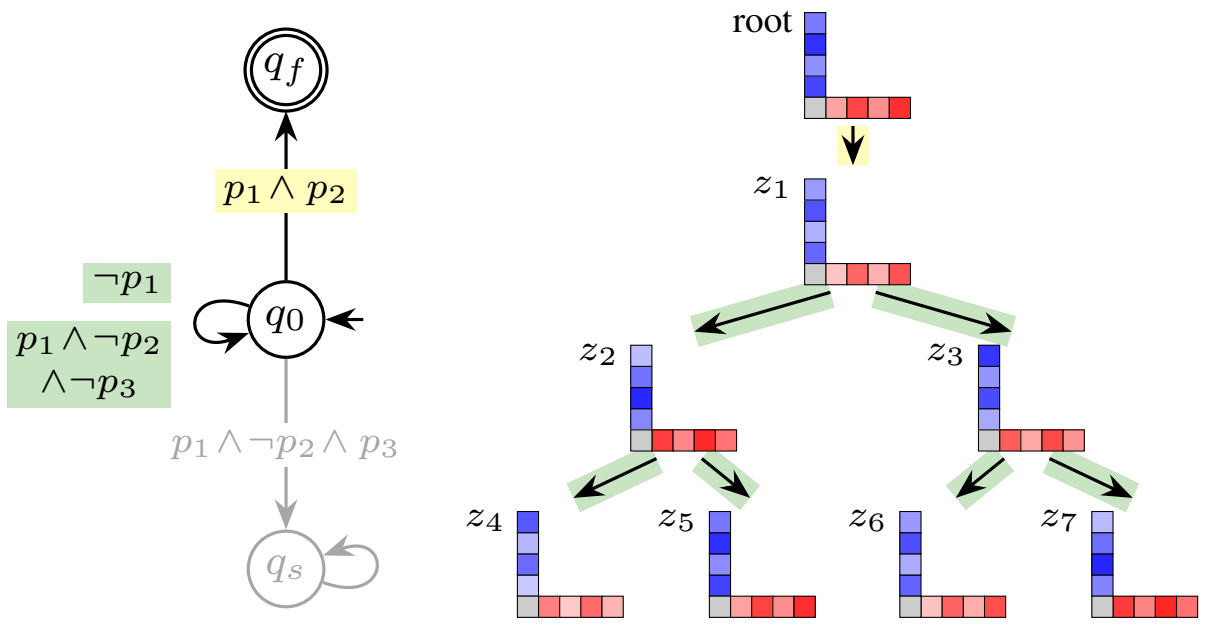

A mode-dependent policy is decoupled, that is, $\boldsymbol{\pi}\in \Pi_{\text{dec}}$, if each agent chooses its input from its own state and the mode, $\pi_q[t](s)=(\pi_{q,1} [t](s_1),\ldots,\pi_{q,N}[t](s_N))$. Alg.~\ref{alg:tree} maps the tree from $\mathcal G_t$ to $\mathcal G_{t+1}$:
\begin{equation} \label{eq:tree_grow}
    \begin{aligned}
        \mathcal G_{t+1} = \mathcal T^{\pi[T-1-t]}(\mathcal G_t),\quad t=0,\ldots,T-1
    \end{aligned}
\end{equation}
with initialization $\mathcal G_0=(\{\text{root}\}, \emptyset, \mathcal{L}_Q(\text{root})=q_f,\valuemapping(\text{root})=\indicator)$. We call the tree computed by \eqref{eq:tree_grow} the RMDP tree. 

\begin{algorithm}[htbp]
\caption{$\mathcal T^{\pi}(\mathcal G)$: RMDP tensor tree value iteration }\label{alg:tree}
\begin{algorithmic}[1]
\Require tree $\mathcal G$; policy $\pi\in\Pi_{\text{dec}}$; pruning threshold $\theta\ge0$
\Ensure tree $\mathcal G$ after one value iteration
\For{each leaf $z$ of $\mathcal G$}\Comment{Grow}
  \For{each $(\alpha,q)$ with $\tau_{\mathcal A}(q,\alpha)=\mathcal L_Q(z)$}
        \State $\bar z\gets$ CreateNewVertex; $\mathcal Z\gets\mathcal Z\cup\{\bar z\}$
    \State $\mathcal E\gets\mathcal E\cup\{(z,\subphi,\bar z)\}$; $\mathcal L_Q(\bar z)\gets q$
  \EndFor
\EndFor
\For{each $(z,\alpha,\bar z)\in\mathcal E$, children first}\Comment{Update}
  \For{$i=1,\dots,N$, with $q=\mathcal L_Q(\bar z)$}
    \State $\eta^i(\bar z)\gets\min_{\xi_{q,i}}\mathbf T^{\pi_{q,i},\xi_{q,i}}_{\alpha_i}(\eta^i(z))$ 
  \EndFor
\EndFor
\For{each leaf $z$ with $\prod_i\max_{s_i}\eta^i(z)(s_i)<\theta$}\Comment{Prune}
  		\State $\mathcal{G}.\mathcal Z \leftarrow \mathcal Z \setminus \{z\}$, $\mathcal{G}.\mathcal{E}\leftarrow \mathcal{E} \setminus \{\text{parent}(z),\subphi,z\}$
\EndFor
\State \Return $\mathcal G$
\end{algorithmic}
\end{algorithm}
\par Fig.~\ref{fig:tree} shows the tensor tree of Example~\ref{example} after three steps, $\mathcal G_3$, without pruning. Each vertex tracks one accepting run of the DFA. Going upwards from any vertex to the root, each edge's Boolean formula triggers one step of this run. For instance, 
\begin{equation}
    \begin{aligned}
        &z_1: q_0\xrightarrow{p_1\wedge p_2}q_f,
        \\
        &z_2:q_0 \xrightarrow{\neg p_1} q_0 \xrightarrow{p_1\wedge p_2} q_f,
        \\
        &z_3: q_0 \xrightarrow{p_1 \wedge \neg p_2 \wedge \neg p_3} q_0 \xrightarrow{p_1 \wedge p_2}q_f.
    \end{aligned} \nonumber
\end{equation}

\begin{theorem}
    [Rank-$1$ tree vertex]\label{theorem:rank1_node}
    \hspace{1mm} For $\boldsymbol{\pi}\in \Pi_{\text{dec}}$ and the trees $\mathcal G_t$ computed by \eqref{eq:tree_grow}, every vertex value is a rank-$1$ tensor, $\valuemapping_t(z)=\bigotimes_{i=1}^N \valuemapping_t^i(z)$ with $\valuemapping_t^i(z)\in \mathbb R^{|\spaceS_i|}$.

    
\end{theorem}

\begin{theorem}
    [RMDP tree as a lower bound] \label{theorem:tree_lower_bound}
    \hspace{1mm} For a horizon $T$, a policy $\{\pi_q[t]|q\in Q, t=0,\ldots, T-1\}\in \Pi_{\text{dec}}$ and trees $\mathcal{G}_t=(\mathcal{Z}_t,\mathcal{E}_t,\mathcal{L}_{Q_t},\valuemapping_t)$ computed by \eqref{eq:tree_grow}, 
    \begin{equation} \label{eq:tree_value_function}
        \begin{aligned}        \underline{\tensorV}_{q,t}^{\boldsymbol{\pi}} \geq \sum_{z\in \mathcal Z_t:\mathcal L_Q(z)=q} \valuemapping_t(z) \quad \text{for all }q\in Q, t\leq T,
        \end{aligned}
    \end{equation}
    where 
    the right-hand side is $\boldsymbol{0}$ if no vertex is labeled $q$.

\end{theorem}Both proofs are by induction over the iterations and are given in the Appendix. With Theorem~\ref{theorem:robust_lb}, evaluating the right-hand side of \eqref{eq:tree_value_function} at $s_0$ and mode $\tau_\DFA(q_0,L(g(s_0)))$ certifies a lower bound for $t=T$ on the satisfaction probability in \eqref{eq:pb_statement}.
\par By Theorem~\ref{theorem:rank1_node}, the right-hand side of \eqref{eq:tree_value_function} is a CPD whose rank-$1$ terms are the vertices with mode $q$. Each vertex stores $\sum_i |\spaceS_i|$ numbers, the tree thus requires $O(|\mathcal Z| \sum_i |\spaceS_i|)$ memory instead of $O(|Q| \prod_i |\spaceS_i|)$, and no value iteration step involves the joint state space or the joint ambiguity set. For $8$ agents with $10^3$ states each, a vertex stores $8\times 10^3$ numbers instead of $ 10^{24}$. The price is the number of vertices $|\mathcal Z|$, which mirrors the number of accepting runs. Longer horizons, or more Boolean formulas per transition, or cycles in the DFA each create more accepting runs, and thus more vertices. Line~12-14 of Alg.\ref{alg:tree} limit the size of the tree via pruning.
\subsection{Policy optimization}
The lower bound of Theorem~\ref{theorem:tree_lower_bound} holds for every decoupled policy, and its tightness depends on the choice of policy. To tighten this bound, we optimize the policy heuristically at each step, after the tree grows and before its values are updated (between line~6 and 7 of Alg.~\ref{alg:tree}). Agent $i$ chooses 
\begin{equation} \label{eq:pol_optimization}
    \begin{aligned}
         \pi_{q,i}(s_i) \!\in\! \arg\max_{u_i\in \spaceU_i} \!\! \sum_{\substack{(z,\subphi,\bar z)\in\mathcal E:\\ \mathcal L_Q(\bar z)=q}} \prod_{j\neq i} c_j(\bar{z}) \min_{\xi_{q,i}} \op_{\subphi_i}^{u_i,\xi_{q,i}} (\valuemapping^i(z))(s_i)
    \end{aligned}
\end{equation}
with $c_j(\bar{z}):= \frac{1}{|\spaceS_j|} \sum_{s_j\in \spaceS_j}  \min_{\xi_{q,j}} \op_{\subphi_j}^{\bar{\pi}_{q,j},\xi_{q,j}} (\valuemapping^j(z))(s_j) \in [0,1]$, where $\bar{\pi}_{q,j}$ is the  policy of agent $j$ from the previous iteration step. Since agent $i$ does not observe the state $s_j$ of agent $j$, it uses $c_j$, the average of agent $j$'s factor over $\spaceS_j$, which weights each vertex by how likely agent $j$ is to satisfy its part of the Boolean formula, $\subphi_j$. Rule \eqref{eq:pol_optimization} maximizes an approximation of the right-hand side of \eqref{eq:tree_value_function}, in which the states of the other agents are averaged out, and their policies are fixed to those of the previous iteration. For a single-agent system, no average is needed, and \eqref{eq:pol_optimization} maximizes the right-hand side of \eqref{eq:tree_value_function} exactly. Each $c_j$ is a scalar computed from agent $j$ alone, and \eqref{eq:pol_optimization} is solved over the input of agent $i$ separately for all $i\in[1,N]$. Hence, no step of the policy optimization involves the joint state space. 

\section{Case studies}
We compare RMDP tree with representatives of the certified approaches to the same problem: SimRel tree \cite{wang2026correct}, SySCoRe \cite{van2023syscore}, and odIMDP \cite{mathiesen2025scalable}\footnote{RMDP tree and SimRel tree \cite{wang2026correct} in Python, SySCoRe \cite{van2023syscore} in MATLAB, odIMDP \cite{mathiesen2025scalable} in Julia. 
}. We do not include compositional approaches, as they do not certify a lower bound for specifications that couple the agents. All methods ran on nodes of the TU/e high-performance cluster with AMD EPYC 9654P or 9474F processors, with 12 threads available each. 
\par Each agent of the continuous-state system is abstracted into a finite RMDP by gridding its state space, adding an absorbing exit cell, from which $\phi$ is counted as violated, and bounding the transition probabilities over each cell, as in \cite{mathiesen2025scalable}. The grids align with the regions of the atomic propositions, such that each proposition holds either on a whole cell or nowhere on it. We apply the policy computed on the RMDP to the cell of the continuous state. On the cells, the continuous-state system then evolves as the RMDP under an adversary that, at each step, chooses the transition distribution of the current continuous state. By construction of the abstraction, this distribution lies in $\Gamma(s,u)$. The proof of Theorem~\ref{theorem:robust_lb} holds for every adversary whose choice lies in $\Gamma(s,u)$ at each step, including this one. Hence, the certified lower bound of Theorem~\ref{theorem:tree_lower_bound} also holds for the continuous-state system.

\subsection{Car parking}
\label{subsec:carpark}
\begin{figure}[htbp]
    \centering
    \includegraphics[width=\linewidth]{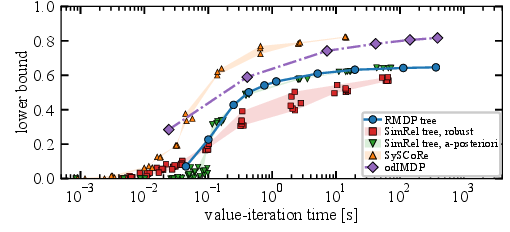}
    \caption{Mean lower bound at $T=20$ over the initial cells outside $p_1\wedge p_2$ and $p_1\wedge p_3$, against the value iteration time (mean of $10$ runs). SimRel trees and SySCoRe: one marker per resolution and $(\epsilon,\delta)$ pair. RMDP tree and odIMDP: one marker per resolution. Finer resolution lies further right. 
    The resolution ranges from $400$ joint cells to $10^7$ for RMDP tree and SySCoRe, $4.1\times 10^7$ for SimRel tree, and $4\times 10^4$ for odIMDP.
    }
    \label{fig:carpar_2d_compare}
\end{figure}
Consider the 2D car parking benchmark of \cite{van2023syscore,mathiesen2025scalable}. Each agent has dynamics $x_i^+ = 0.9x_i+0.7u_i+w_i$, $w_i \sim \mathcal N(0,1)$, $\spaceX_i=[-10,10]$, $\spaceU_i=[-1,1]$, $y_i=x_i$, $i=1,2$. The specification is $\phi:= \neg(p_1 \wedge p_3) \until (p_1\wedge p_2)$ where $p_1$ is true iff $y_1\in[4,10]$, $p_2$ iff $y_2\in [-4,0]$, and $p_3$ iff $y_2\in [0,4]$. 
\par We grid each $\spaceX_i$ with $20$ cells up to $3200$ for RMDP tree and SySCoRe, $6400$ for SimRel tree and $200$ for odIMDP\footnote{For all methods and grids, the input per agent is $\{-1,0,1\}$.}. Fig.~\ref{fig:carpar_2d_compare} shows 
the results for $T=20$.
The reported value-iteration time excludes the abstraction.
\par This benchmark favours the monolithic approaches: SySCoRe and odIMDP certify up to $0.82$, while RMDP tree certifies at most $0.65$, even at $3200$ cells per agent. The policies of monolithic abstraction approaches choose each input from both coordinates of the car, whereas the decoupled policy of the tree approaches uses only the coordinate of its own agent. However, the value-iteration time of odIMDP grows faster with the abstraction resolution. Doubling the cells per agent multiplies its time by $17$ to $20$, against $2$ to $6$ for RMDP tree and $3$ to $5$ for SySCoRe.
\par At the same abstraction resolution, RMDP tree certifies a $1.2$ to $7$ times higher lower bound than SimRel tree. To certify the same lower bound, RMDP tree also needs far less time: $0.50$ in $0.43$ s, $5$ times faster, and $0.61$ in $5.2$ s, $12$ times faster.

\subsection{Ordered reach-avoid}
\begin{figure}[htbp]
\centering
\begin{tikzpicture}[>=Stealth, shorten >=1pt, auto, semithick,
    every state/.style={minimum size=6mm, inner sep=0pt},
    every node/.style={font=\footnotesize}]
  \node[state, initial, initial text={}] (q0) at (0,0)     {$q_0$};
  \node[state]                           (q1) at (3.0,0)   {$q_1$};
  \node[state, accepting]                (qf) at (5.4,0)   {$q_f$};
  \node[state]                           (qs) at (3.0,-1.2) {$q_s$};
  \path[->] (q0) edge[loop, out=125, in=55, looseness=7] node {\scriptsize$\neg p_2\wedge\neg p_4$} (q0);
  \path[->] (q1) edge[loop, out=125, in=55, looseness=7] node {\scriptsize$\neg p_2$} (q1);
  \path[->] (q0) edge node {\scriptsize$\neg p_2\wedge p_3\wedge p_4$} (q1);
  \path[->] (q1) edge node {\scriptsize$p_1\wedge p_2$} (qf);
  \path[->] (q0) edge[bend right=15] node[swap,align=center] {\scriptsize$p_2$,\\[-1pt] \scriptsize$\neg p_2\wedge\neg p_3\wedge p_4$} (qs);
  \path[->] (q1) edge node {\scriptsize$\neg p_1\wedge p_2$} (qs);
  \path[->] (qs) edge[loop, out=35, in=-35, looseness=7] (qs);
\end{tikzpicture}
\caption{DFA of the ordered reach-avoid specification.}
\label{fig:drones_dfa}
\end{figure}
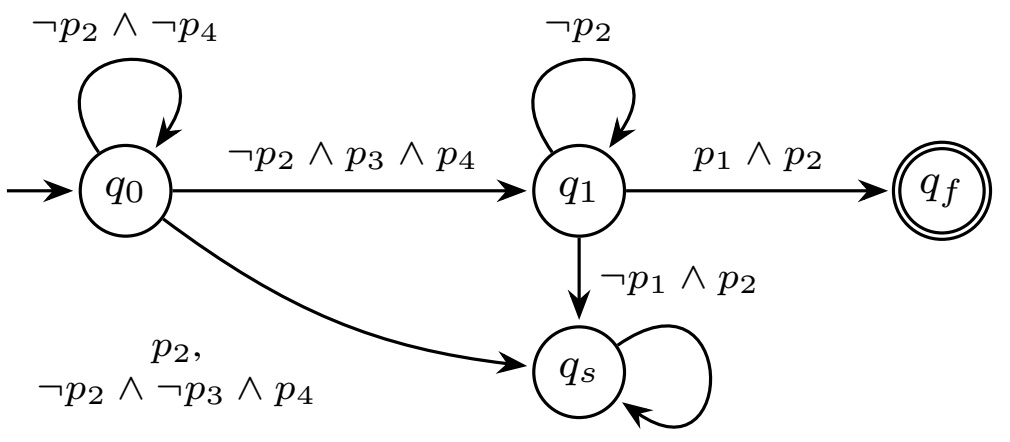

We extend the car parking benchmark to two drones that reach designated zones in order. Each drone has two coordinates, and each coordinate is modelled as one agent. Each agent has the dynamics, state space, and inputs of Sec.~\ref{subsec:carpark}. Drone 1 has outputs $y_1,y_2$, and drone 2 has $y_3,y_4$. $p_1$ is true iff $y_1\in [-10,-4]$, $p_2$ iff $y_2\in [2,10]$, $p_3$ iff $y_3\in [4,10]$, and $p_4$ iff $y_4\in [2,10]$. The specification is $\phi:= (\neg p_2 \wedge \neg p_4) \until (  \neg p_2 \wedge p_3 \wedge p_4 \wedge (\neg p_2 \until (p_1 \wedge p_2)))$, with its DFA in Fig.~\ref{fig:drones_dfa}. Drone 2 must reach $[4,10] \times [2,10]$ before drone 1 reaches $[-10,-4]\times [2,10]$. Before reaching its zone, neither drone may enter $[-4,4]\times [2,10]$.

\par SySCoRe grids only 2D models and therefore does not apply. 
Tab.~\ref{tab:drones} reports the results for $T=30$.
\begin{table}[t]
\centering\footnotesize\setlength{\tabcolsep}{4pt}
\caption{Computation and satisfaction probabilities for the ordered reach-avoid with $T=30$. Cells counts the joint cells, time is the mean value-iteration time over $10$ runs. Bound denotes the mean lower bound over the initial cells in $y\in[-10,2]$, and emp. the empirical success rate in 2000 Monte Carlo runs.}
\label{tab:drones}
\begin{tabular}{@{}cccccc@{}}\toprule
method & cells & memory (MB) & time (s) & bound & emp.\\\midrule
RMDP tree  & $4.1\!\!\times\!\! 10^{7}$  & 2.2 & 3.3  & 0.146 & 47\%\\
            & $6.6\!\!\times\!\! 10^{8}$  & 6.9 & 8.2  & 0.287 & 47\%\\
            & $1.0\!\!\times\!\! 10^{10}$ & 24  & 34.7 & 0.372 & 46\%\\
            & $1.7\!\!\times\!\! 10^{11}$ & 87  & 130  & 0.414 & 45\%\\\midrule
SimRel tree, & $4.1\!\!\times\!\! 10^{7}$  & 0.6 & 0.05 & $3\!\!\times\!\! 10^{-4}$ & 2\%\\
            robust& $6.6\!\!\times\!\! 10^{8}$  & 2.6 & 0.14  & 0.002 & 10\%\\
            & $1.0\!\!\times\!\! 10^{10}$ & 11  & 0.93  & 0.019 & 27\%\\
            & $1.7\!\!\times\!\! 10^{11}$ & 45  & 4.3 & 0.072 & 44\%\\\midrule
SimRel tree, & $4.1\!\!\times\!\! 10^{7}$  & 0.8 & 0.13 & 0 & 31\%\\
            a-posteriori& $6.6\!\!\times\!\! 10^{8}$  & 4.2 & 0.70  & 0 & 42\%\\
            & $1.0\!\!\times\!\! 10^{10}$ & 14  & 1.4  & 0 & 43\%\\
            & $1.7\!\!\times\!\! 10^{11}$ & 47  & 3.8 & 0 & 44\%\\\midrule
SySCoRe     & \multicolumn{5}{c}{does not apply (SySCoRe grids only 2D models)}\\\midrule
odIMDP      & $10^{4}$           & 518 & 805 & $7\!\!\times\!\! 10^{-5}$ & 37\%\\
            & $1.6\!\!\times\!\! 10^{5}$  & 17000 & $1.7\!\!\times\! \! 10^5$$^\dagger$ & 0.027 & 56\%\\\bottomrule
\end{tabular}
\vspace{2pt}
\parbox{\columnwidth}{\footnotesize $^\dagger$Single run, since one run takes about 46.8 hours. }
\vspace{-4mm}
\end{table}
Over $2000$ Monte Carlo runs at the finest resolution of each method, it is shown in Tab.~\ref{tab:drones} that all three methods synthesize controllers at a comparable success rate. 
\par With four agents, a fine abstraction of the multi-agent system becomes computationally hard, which prevents the monolithic approaches from exploiting their advantages as in Sec.~\ref{subsec:carpark}. odIMDP stores transition bounds for every joint cell, so its memory grows exponentially with the number of agents: $518$ MB for $10$ cells per agent and $17$ GB for $20$ cells per agent. At $20$ cells, it certifies $0.027$, below the $0.146$ of RMDP tree at its coarsest resolution. RMDP tree stores transition bounds per agent, so its memory grows linearly with the number of agents, and needs $87$ MB at $640$ cells per agent. 
SimRel tree is more conservative: RMDP tree at its coarsest resolution certifies a lower bound twice as high as SimRel tree at its finest resolution, with $20$ times less memory.

\section{Conclusion}
We presented RMDP tree, a method for controller synthesis of multi-agent RMDPs under temporal logic specifications that couple the agents. We showed that for agents that move and are disturbed independently of each other, a lower bound on the satisfaction probability can be stored as a tensor tree, whose memory grows linearly with the number of agents instead of exponentially. Compared with SimRel tree, which also uses a tensor tree and needs less time and memory per resolution, RMDP tree certifies tighter lower bounds at the same resolution. Our case studies support these findings.
\par Our method has limitations. First, it applies only to agents that are decoupled in their dynamics, uncertainty, and atomic propositions. Coupled constraints, such as collision avoidance, can be encoded only conservatively through per-agent region labels, and handling them efficiently is an important future direction. Second, each agent chooses its input from its own state and the DFA state, which loosens the bound when the task benefits from reacting to the other agents. Extending the automaton with additional modes would let the policies react to more of the joint behaviour. Finally, the tree grows with the number of accepting runs, which can grow exponentially with the horizon for cyclic specifications. Pruning keeps the tree small but loosens the bound. Compressing the tree without loosening the bound is a natural next step. 

\vspace{-1.5ex}
\bibliographystyle{IEEEtran}
\bibliography{references}

\vspace{-1.5ex}
\appendices
\section*{Appendix}
\begin{proof}[Proof of Theorem~\ref{theorem:robust_lb}]
\renewcommand{\qedsymbol}{}
We arbitrarily choose a history-dependent adversary $\boldsymbol{\xi} \in \Xi$. For a history $\history[t]=s[0],u[0],\ldots,u[t-1],s[t]$ that ends in $(s,q)$, we denote by $W_{q,T-t}(\history[t])$ the probability that the word is accepted within the remaining $T-t$ steps, given that $\mdpM \times \boldsymbol{\pi}\times \boldsymbol{\xi}$ has produced $\history[t]$. $W_{q,T-t}$ is iteratively computed similar to \eqref{eq:vi_fixed_pol_adv}: with $u=\pi_q[t](s)$ and $\gamma=\xi[t](\history[t],u)\in \Gamma(s,u)$,
\begin{equation} \label{eq:defW}
    \begin{aligned}
        &W_{q,T-t}(\history[t])
        \\
        &=\indicator_{q_f}(q)+\indicator_{Q\setminus \{q_f\}} (q)\sum_{s'\in\spaceS} \gamma(s')W_{q',T-t-1} (\history[t],u,s')
    \end{aligned}
\end{equation}
    with $q'=\tau_\DFA(q,L(g(s')))$, and $W_{q,0}(\history[T])=\indicator_{q_f}(q)$. As in Proposition~\ref{prop:spec_satisfaction}, $\mathbb P_{\mdpM\times \boldsymbol{\pi} \times \boldsymbol{\xi}}^{s_0}(\word_{[0,T]}\models \phi)=W_{\bar{q}_0,T}(s_0)$. 
    We prove that for all $t\in [0,T]$ and all histories $\history[t]$ ending in $(s,q)$,
    \begin{tightdisplay}
    \begin{equation} \label{eq:WgeqVlb}
        \begin{aligned}
        W_{q,T-t}(\history[t]) \geq \underline{\tensorV}_{q,T-t}^{\boldsymbol{\pi}}(s).
        \end{aligned}
    \end{equation}
    \end{tightdisplay}
    For $t=T$, both sides are $\indicator_{q_f}(q)$ by the initialization of \eqref{eq:defW} and \eqref{eq:robust_valuefunc}. We then prove that if \eqref{eq:WgeqVlb} holds for $t+1$, then it also holds for $t$. For $q=q_f$, both sides are $\indicator$. For $q\in Q\setminus \{q_f\}$,
    \begin{tightdisplay}
    \begin{equation}
        \begin{aligned}
            &W_{q,T-t}(\history[t]) 
            \\
            & =\sum_{s'\in \spaceS} \gamma(s') W_{q',T-1-t}(\history[t],u,s')
            \\
            & \geq \sum_{s'\in \spaceS} \gamma(s')\underline{\tensorV}_{q',T-1-t}^{\boldsymbol{\pi}}(s')
            \\
            & = \sum_{(\subphi,q') \in \text{Succ}(q)} \sum_{s'\in \spaceS} \indicator[L(g(s'))\models \subphi] \gamma(s')\underline{\tensorV}_{q',T-1-t}^{\boldsymbol{\pi}}(s')
            \\
            & \geq \min_{\gamma'\in \Gamma(s,u)} \sum_{(\subphi,q') \in \text{Succ}(q)} \sum_{s'\in \spaceS} \indicator[L(g(s'))\models \subphi] \gamma'(s')\underline{\tensorV}_{q',T-1-t}^{\boldsymbol{\pi}}(s')
            \\
            & =\underline{\tensorV}_{q,T-t}^{\boldsymbol{\pi}}(s),
        \end{aligned}
    \end{equation}
    \end{tightdisplay}
    where the first inequality is \eqref{eq:WgeqVlb} for $t+1$, and the last equality is \eqref{eq:robust_valuefunc}, whose minimum at $t$ is over $\xi_q[t](s,u)\in \Gamma(s,u)$ with $u=\pi_q[t](s)$. At $t=0$, $\history[0]=s_0$ with $q=\bar{q}_0$, and $W_{\bar{q}_0,T}(s_0)=\mathbb P_{\mdpM\times \boldsymbol{\pi} \times \boldsymbol{\xi}} ^{s_0}(\word_{[0,T]}\models \phi) \geq \underline{\tensorV}_{\bar{q}_0,T}^{\boldsymbol{\pi}}(s_0)$.
\end{proof}

\begin{proof}[Proof of Theorem~\ref{theorem:rank1_node}]
\renewcommand{\qedsymbol}{}
    The proof follows that of \cite[Theorem~3]{wang2025unraveling}. Let $\valuemapping(z)=\bigotimes_i \mathrm v^i$ with nonnegative $\mathrm v^i$. With $\subphi=\subphi_1\wedge\ldots\wedge\subphi_N$, the sum in \eqref{eq:op_general} applied to $\valuemapping(z)$ is the product over $i$ of the sums in \eqref{eq:op_general} for agent~$i$, and the $i$-th factor depends only on $\gamma_i$. By the definition of $\Gamma(s,u)$, every $\gamma\in\Gamma(s,u)$ is a product $\gamma=\prod_i\gamma_i$ with $\gamma_i\in\Gamma_i(s_i,u_i)$, and each $\gamma_i$ can be chosen independently. The minimum of a product of nonnegative factors is then the product of their minima:$
    \min_{\xi_q}\op_{\subphi}^{\pi_q,\xi_q}(\valuemapping(z))=\bigotimes_{i=1}^N\min_{\xi_{q,i}}\op_{\subphi_i}^{\pi_{q,i},\xi_{q,i}}(\mathrm v^i),
    $
    which is line~9 of Alg.~\ref{alg:tree}. Induction over $t$ from $\valuemapping(\text{root})=\indicator$ completes the proof.
\end{proof}

\begin{proof}[Proof of Theorem~\ref{theorem:tree_lower_bound}]
    \enlargethispage{2\baselineskip}
    \renewcommand{\qedsymbol}{}For $t=0$, $\mathcal G_0$ contains only the root. Both sides of \eqref{eq:tree_value_function} are $\boldsymbol{1}$ for $q=q_f$ and $\boldsymbol{0}$ otherwise. We then prove that if \eqref{eq:tree_value_function} holds for $t=k$, then it also holds for $t=k+1$. For $q=q_f$, Alg.~\ref{alg:tree} creates no $q_f$-labeled vertex beyond the root node, so both sides of \eqref{eq:tree_value_function} are $\boldsymbol{1}$ for all $t\in [ 0,T ]$. For $q\in Q\setminus \{q_f\}$, we start by expressing the left-hand side of \eqref{eq:tree_value_function} for $t=k+1$ based on \eqref{eq:robust_valuefunc}, with $\pi_q=\pi_q[T-1-k]$ and $\xi_q=\xi_q[T-1-k]$:
    \begin{equation}
        \begin{aligned}
            \underline{V}_{q,k+1}^{\boldsymbol{\pi}}&=\min_{\xi_q}\sum_{(\subphi,q')\in {\small \text{Succ}}(q) } \op_{\subphi}^{\pi_q,\xi_q}(\underline{\tensorV}_{q',k}^{\boldsymbol{\pi}})
                    \\         
        &  \downarrow \text{\eqref{eq:tree_value_function} holds for $t=k$, and \eqref{eq:op_general} is monotone} \notag
        \\
        &\geq \min_{\xi_q}\sum_{(\subphi,q')\in {\small \text{Succ}}(q) } \op_{\subphi}^{\pi_q,\xi_q} \bigl(  \sum_{z \in \mathcal{L}_{Q_k}^{-1}(q') } \valuemapping_k(z) \bigr)
        \\
        &=\min_{\xi_q}\sum_{(\subphi,q')\in {\small \text{Succ}}(q) } \sum_{z \in \mathcal{L}_{Q_k}^{-1}(q') } \op_{\subphi}^{\pi_q,\xi_q} \bigl( \valuemapping_k(z) \bigr)
        \\
&\downarrow\ \text{\small\itshape vertices removed by pruning only }\notag\\
&\quad\ \text{\small\itshape  drop nonnegative terms \cite[Lemma~3]{wang2025unraveling}}\notag\\
        &\geq\min_{\xi_q} \sum_{\substack{(z,\subphi,\bar z)\in\mathcal{E}_{k+1}\\ \mathcal{L}_{Q_{k+1}}(\bar z)=q}} \op_{\subphi}^{\pi_q,\xi_q}(\valuemapping_k(z))
        \\
        &=\min_{\xi_q} \sum_{\substack{(z,\subphi,\bar z)\in\mathcal{E}_{k+1}\\ \mathcal{L}_{Q_{k+1}}(\bar z)=q}} \bigotimes_{i=1}^N \op_{\subphi_i}^{\pi_{q,i},\xi_{q,i}} \bigl( \valuemapping_k^i(z) \bigr)
         \\   
         &\downarrow\ \text{\small\itshape the minimum of a sum is at least the sum of the minima}\notag\\
        &\geq \sum_{\substack{(z,\subphi,\bar z)\in\mathcal{E}_{k+1}\\ \mathcal{L}_{Q_{k+1}}(\bar z)=q}} \min_{\xi_{q}} \bigotimes_{i=1}^N \op_{\subphi_i}^{\pi_{q,i},\xi_{q,i}} \bigl( \valuemapping_k^i(z) \bigr)
        \\
     &\downarrow\ \text{\small\itshape Theorem~\ref{theorem:rank1_node}}\notag\\
        &=\sum_{\substack{(z,\subphi,\bar z)\in\mathcal{E}_{k+1}\\ \mathcal{L}_{Q_{k+1}}(\bar z)=q}}  \bigotimes_{i=1}^N \min_{\xi_{q,i}}\op_{\subphi_i}^{\pi_{q,i},\xi_{q,i}} \bigl( \valuemapping_k^i(z) \bigr)
        \\
&\downarrow\ \text{\small\itshape line~9 of Alg.~\ref{alg:tree}}\notag\\
        &=\sum_{\substack{(z,\subphi,\bar z)\in\mathcal{E}_{k+1}\\ \mathcal{L}_{Q_{k+1}}(\bar z)=q}}  \valuemapping_{k+1}(\bar{z})= \sum_{\bar{z}\in\mathcal{L}_{Q_{k+1}}^{-1} (q)} \valuemapping_{k+1}(\bar{z}).
        \end{aligned}
    \end{equation}
\end{proof}
\vspace{-4mm}
\end{document}